# Synthesis, crystal and electronic structures, and second harmonic generation of La$_4$Ge$_3$S$_{12}$

Hiroya Ohtsuki,[a] Suguru Nakata,[a] Yu Yamane,[a] Ryunosuke Takahashi,[a] Koichi Kusakabe,[a] and Hiroki Wadati*[a,b]

**Abstract:** The crystal structure of La$_4$Ge$_3$S$_{12}$ has been known to be noncentrosymmetric for almost four decades. This characteristic inversion symmetry breaking suggests the presence of nonlinear optical properties. Yet only recently have nonlinear optical phenomena such as second harmonic generation (SHG) been reported in this material. In this study, we synthesized La$_4$Ge$_3$S$_{12}$ using the direct reaction method and characterized the composition, crystal structure, and electronic structure using electron probe microanalysis, powder and single-crystal X-ray diffraction, and X-ray photoelectron spectroscopy. The experimentally measured electronic structure is in line with that obtained using first-principles calculations. In addition, we observed the nonlinear optical properties of La$_4$Ge$_3$S$_{12}$ in response to an ultrashort infrared pulsed laser. We found that the intensity of the SHG depends quadratically on the intensity of the incident light, mirroring the intrinsic nature of nonlinear optics.

## Introduction

Second harmonic generation (SHG) is a powerful method that has significantly extended the energy range of lasers opening the possibility of their use in a variety of applications such as telecommunications, atmospheric sensing, medical applications, and biological imaging [1]. A key aspect of SHG is the second-order electric nonlinear optical (NLO) response, which is particularly sensitive to the space group of the material. This response can occur in crystals with broken spatial inversion symmetry, and therefore it is important to investigate the NLO response in these materials since this could help further extend the energy range of lasers.

Recently, inorganic sulfides with spatial inversion symmetry breaking have attracted increasing attention as a NLO material. Compared to an O$^{2-}$ ion, a S$^{2-}$ ion possesses a larger ionic radius of 1.84 Å (coordination number 6) and smaller electronegativity of 2.58 (Pauling's scale). The chemical properties of sulfides allow them to crystallize in characteristic crystal structures with chirality and/or polarity. In fact, the ternary sulfide AgGaS$_2$ (AGS) with tetragonal space group *I*-42*d* shows a large SHG response and has been used as an infrared light (IR)-NLO material. Recently, quaternary sulfides (RE)$_6$B$_x$C$_y$S$_{14}$ (RE: rare earth, B and C: various elements such that the sum of the valence electrons = 10) in the chiral polar space group *P*6$_3$ ($C_6^6$, No.173) have also been studied [2–11].

In another family of similar materials, Y$_4$Si$_3$S$_{12}$ [12], La$_4$Ge$_3$S$_{12}$, and the Se substituted system La$_4$Ge$_3$Se$_x$S$_{12-x}$ [13, 14] all with the polar space group *R*3*c* ($C_{3v}^6$, No.161), the prerequisite for NLO properties can be easily fulfilled. For Y$_4$Si$_3$S$_{12}$, density functional theory (DFT) calculation predicts a moderate SHG response of $\chi^{111}$ = 15.303 pm/V [12]. Experimentally, the SHG intensity and the laser damage threshold (LDT) have been studied for La$_4$Ge$_3$Se$_x$S$_{12-x}$ with either a Q-switched laser (2.09 μm, 3 Hz, 50 ns) or a pulsed YAG laser (1.06 μm, 10 ns, 10 Hz). By optimizing the bandgap by Se substitution, La$_4$Ge$_3$Se$_4$S$_8$ shows the highest SHG response of 9.4 times that of AGS, and the LDT is 2.4 times that of AGS. Despite the NLO properties of ns-range pulsed lasers being well established, those for ultrashort (femtosecond) pulsed lasers remain relatively unknown.

With this in mind, we prepared polycrystalline samples of La$_4$Ge$_3$S$_{12}$ using a direct reaction method and investigated their physical properties and NLO characteristics. We performed electron probe microanalysis, powder and single-crystal X-ray diffraction (XRD), and X-ray photoelectron spectroscopy (XPS) to characterize the composition, crystal structure, and electronic structure. Moreover, we investigated the SHG of La$_4$Ge$_3$S$_{12}$. We found that the SHG intensity was quadratically proportional to the intensity of the incident ultrashort pulsed laser and about half of that of a powdered KH$_2$PO$_4$ (KDP) sample.

## Results and Discussion

### Crystal structure and chemical composition

The crystallographic and structure refinement results of La$_4$Ge$_3$S$_{12}$, as derived by single-crystal XRD, are summarized in Table 1. Atomic coordinates and equivalent isotropic displacement parameters are shown in Table 2, and anisotropic displacement parameters are displayed in Table 3. Further information, in CIF format, has been deposited at the Cambridge Crystallographic Data Centre under reference number CCDC-2421046.

The space group of La$_4$Ge$_3$S$_{12}$ is trigonal *R*3*c*. Here, we chose hexagonal axes to describe the lattice parameters and atomic coordinates of this material. The crystal structure is displayed in Fig. 1 (a) and (b), which is isostructural to other compounds of La$_4$Ge$_3$Se$_x$S$_{12-x}$ [14]. The face-sharing La2S$_6$ trigonal prisms form a one-dimensional chain along the c-axis via parallel stacking of the prisms. Figure 1(c) shows the coordination polyhedra of the atoms with the selected bond distances and angles. The La2 atom is slightly off-centered in the trigonal prism, evidenced by the distances of La2-upper S (2.920(2) Å) and La2-lower S (2.909(2) Å). GeS$_4$ tetrahedra reside at a general position located next to the La2S$_6$ trigonal prisms by sharing each corner. The La1S$_7$ polyhedra, with a monocapped trigonal prismatic geometry, are located in a different general position, connecting the GeS$_4$ tetrahedra and another one-dimensional chain. As highlighted in Fig. 1 (b) via a red square, reflecting the noncentrosymmetric space group *R*3*c*, the La1S$_7$ and GeS$_4$ polyhedra located in their respective positions give rise to electric polarity along the c-axis, which subsequently induces the NLO properties in this material.

**Table 1.** Crystallographic and structure refinement data for La$_4$Ge$_3$S$_{12}$.

| Formula | La$_4$Ge$_3$S$_{12}$ |
| --- | --- |
| Formula weight | 1158.13 |
| Crystal size | 0.300 x 0.090 x 0.040 mm$^3$ |
| Color | Orange |

[a] Hiroya Ohtsuki, Suguru Nakata, Yu Yamane, Ryunosuke Takahashi, Koichi Kusakabe, and Hiroki Wadati
Department of Material Science, Graduate School of Science, University of Hyogo
3-2-1 Kouto, Kamigori-cho, Ako-gun, Hyogo 678-1297, Japan
E-mail: wadati@sci.u-hyogo.ac.jp
[b] Hiroki Wadati
Institute of Laser Engineering, Osaka University
2–6 Yamadaoka,Suita,Osaka565-0871,Japan

| Habit | Needle |
|---|---|
| Temperature | 296(2) K |
| Wavelength | 0.71073 Å |
| Crystal system | Trigonal |
| Space group | $R3c$: H (No. 161) |
| Unit cell dimensions | $a$ = 19.4691(14) Å |
| | $c$ = 8.1005(3) Å |
| Volume | 2659.1(4) Å$^3$ |
| $Z$ | 6 |
| Density (calculated) | 4.339 g/cm$^3$ |
| Absorption coefficient | 15.823 mm$^{-1}$ |
| F(000) | 3096 |
| Reflections collected | 1723 |
| Independent reflections | 1723 [$R$(int) = 0.0187] |
| Completeness | 0.998 |
| Absorption correction | Multi-scan |
| Goodness-of-fit on $F^2$ | 0.990 |
| Flack x parameter | 0.018(9) |
| Final $R$ indices [$I$>2sigma(I)] | $R$1 = 0.0186, $wR$2 = 0.0425 |
| $R$ indices (all data) | $R$1 = 0.0190, $wR$2 = 0.0426 |
| Largest diff. peak and hole | 1.189 and -0.948 e.Å$^{-3}$ |

**Table 2.** Atomic coordinates and equivalent isotropic displacement parameters (Å$^2$) for La$_4$Ge$_3$S$_{12}$. $U_{eq}$ is defined as one third of the trace of the orthogonalized $U^{ij}$ tensor.

| | Wyckoff | x | Y | z | $U_{eq}$(Å$^2$) |
|---|---|---|---|---|---|
| La1 | 18b | 0.5612(1) | 0.6639(1) | 0.8701(1) | 0.015(1) |
| La2 | 6a | 0 | 0 | 0 | 0.012(1) |
| Ge1 | 18b | 0.5336(1) | 0.8539(1) | 0.8199(1) | 0.010(1) |
| S1 | 18b | 0.4577(1) | 0.7306(1) | 0.9177(2) | 0.012(1) |
| S2 | 18b | 0.5571(1) | 0.5112(1) | 0.8294(2) | 0.014(1) |
| S3 | 18b | 0.4190(1) | 0.5519(1) | 0.6647(2) | 0.014(1) |
| S4 | 18b | 0.6071(1) | 0.6701(1) | 0.5153(2) | 0.015(1) |

**Table 3.** Anisotropic displacement parameters (Å$^2$) for La$_4$Ge$_3$S$_{12}$. The anisotropic displacement factor exponent takes the form: $-2\pi^2[h^2a^{*2}U^{11} + ... + 2hka^*b^*U^{12}]$

| | $U^{11}$ | $U^{22}$ | $U^{33}$ | $U^{23}$ | $U^{13}$ | $U^{12}$ |
|---|---|---|---|---|---|---|
| La(1) | 0.010(1) | 0.018(1) | 0.016(1) | -0.007(1) | 0.000(1) | 0.006(1) |
| La(2) | 0.015(1) | 0.015(1) | 0.007(1) | 0.000 | 0.000 | 0.008(1) |
| Ge(1) | 0.009(1) | 0.010(1) | 0.010(1) | -0.001(1) | -0.001(1) | 0.004(1) |
| S(1) | 0.011(1) | 0.011(1) | 0.013(1) | 0.001(1) | 0.001(1) | 0.005(1) |
| S(2) | 0.013(1) | 0.016(1) | 0.014(1) | -0.002(1) | -0.003(1) | 0.009(1) |
| S(3) | 0.011(1) | 0.014(1) | 0.012(1) | 0.000(1) | -0.001(1) | 0.002(1) |
| S(4) | 0.020(1) | 0.019(1) | 0.012(1) | -0.002(1) | -0.004(1) | 0.014(1) |

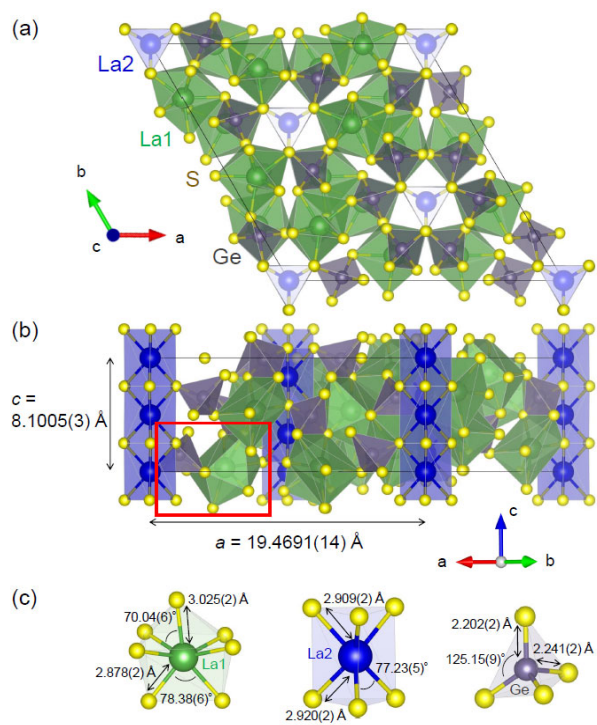

**Figure 1.** Crystal structure of La$_4$Ge$_3$S$_{12}$ viewed along (a) the $c$-direction and (b) the $a^*$-direction, and (c) coordination polyhedra of the atoms. Blue, green, gray, and yellow spheres represent La1, La2, Ge, S atoms, respectively. The La1S7 and GeS4 polyhedra highlighted by a red square break the inversion symmetry. This structure was drawn using VESTA [21].

Fig. 2 shows the powder XRD patterns of a sample of La$_4$Ge$_3$S$_{12}$, which was optimized for SHG measurements. All the peaks in the experimental spectrum (red solid line) can be explained by the simulated ones (black solid line) for the trigonal La$_4$Ge$_3$S$_{12}$-type structure [13, 14], indicating the singe phase of the measured sample. Rietveld refinement confirms the lattice parameters of this sample to be $a$ = 19.471(1) Å and $c$ = 8.1048(5) Å. These values are not only the same as those measured via single-crystal XRD, but are also similar to those previously reported for the same system: $a$ = 19.40 Å and $c$ = 8.10 Å by Mazurier and Etienne [13], and $a$ = 19.427 Å and $c$ = 8.075 Å by Cicirello et al. [14].

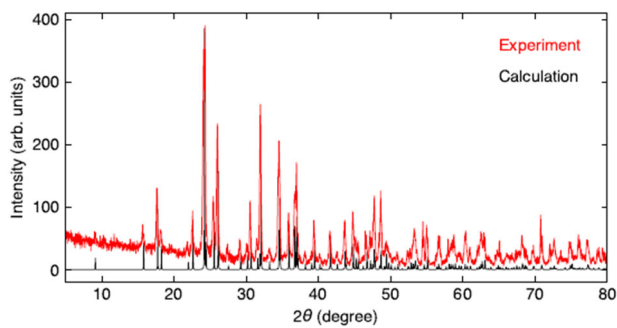

**Figure 2.** Powder X-ray diffraction pattern (red) of $La_4Ge_3S_{12}$ collected at room temperature in comparison with a (black) simulation based on the lattice parameters reported in Ref. [14].

Fig. 3 displays the energy dispersive X-ray spectrum of a $La_4Ge_3S_{12}$ sample. We obtained the atomic ratio La : Ge : S = 21.56 : 15.49 : 62.95 from this spectrum, which shows remarkable agreement with the ideal ratio of La : Ge : S = 4 : 3 : 12 = 21.1 : 15.8 : 63.2.

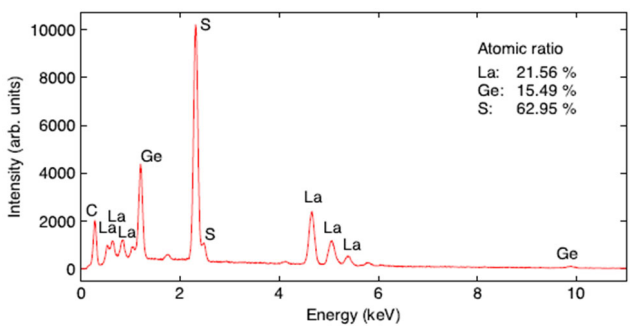

**Figure 3.** An energy dispersive X-ray spectrum of a $La_4G_3S_{12}$ sample collected at room temperature. The labels above each peak indicate elements specific to that peak.

### Electronic structure

Fig. 4 (a-d) shows the core-level XPS spectra, indicating that the La and S core-level peaks are split. The La core-level $3d_{5/2}$ peak can be deconvoluted into two peaks whose energy positions are 835.35 and 838.35 eV. Since the $3d_{5/2}$ peak of $LaRuO_3$, a compound with trivalent La ions, is also split into two components [22], the peak splitting of $La_4Ge_3S_{12}$ is similar to that seen in $LaRuO_3$, suggesting that La is also trivalent in this material. The Ge 3$d$ core-level peak of $La_4Ge_3S_{12}$ is at 31.25 eV, whereas those of tetravalent $GeS_2$ and $GeO_2$ are at 30-31 and 32-33 eV [23], respectively. The peak position in this study is close to the values of $GeS_2$. We therefore argue that the valence states of our samples are the following: $(La^{3+})_4(Ge^{4+})_3(S^{2-})_{12}$.

Fig. 5 (a) displays the valence-band narrow scan from a XPS measurement compared with a core-levels density of states (DOS) calculation. The DOS calculations were based on the crystal structure data from Ref. [14]. The result of the calculated DOS in Fig. 5 (b) is broadened with an energy-dependent Lorentzian (FWHM = $0.2|E - E_F|$ eV) to account for the lifetime broadening of the photohole. Overall, the peak positions of the measured valence-band spectrum are in good agreement with those in the calculated spectrum. Comparing Fig. 5 (a) with Fig. 5 (b), the valence band features within the energy range of 0-6 eV are primarily derived from S 3$p$ states, while those at ~ 8 eV originate mainly from S 3$p$ and Ge 4$sp$ states, and those at ~ 13 eV from S 3$s$ states.

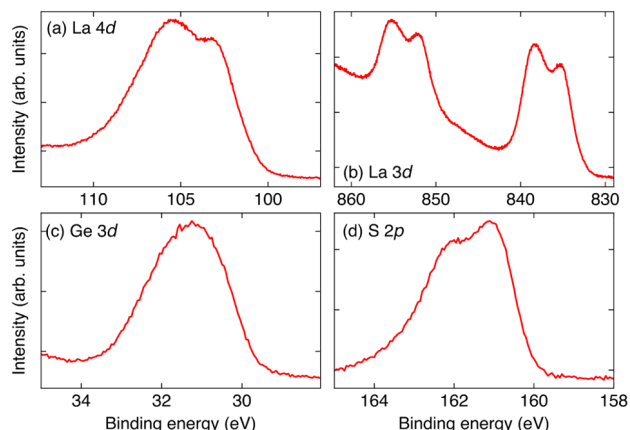

**Figure 4.** (a-d) XPS core-level spectra of $La_4Ge_3S_{12}$ collected at room temperature ($h\nu$ = 1486.6 eV). (a) La 4$d$, (b) La 3$d$, (c) Ge 3$d$, and (d) S 2$p$.

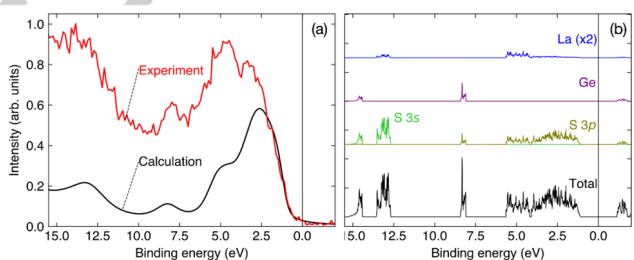

**Figure 5.** (a) A XPS valence-band spectrum of $La_4Ge_3S_{12}$ collected at room temperature (red) in comparison to a first-principles density of states calculation (black). To account for the natural broadening of the experimental curve, the theoretical curve is also broadened. (b) Total and partial density of states of $La_4Ge_3S_{12}$ obtained from first-principles calculations. Note that the partial DOS are vertically shifted, and the intensity of La DOS is multiplied by two for visual clarity.

### Second harmonic generation

Fig. 6 (a) shows the SHG spectra of $La_4Ge_3S_{12}$ (515 nm) in response to an incident ultrashort pulsed laser (1030 nm) with various laser intensities. The relationship between the intensity of the second harmonic (SH) intensity and the laser intensity is clearly nonlinear. Looking at this NLO property in detail, we find that the SH response is easily fitted using a quadratic function, a characteristic behavior of SHG (Fig. 6 (b)). To understand this SH response quantitatively, we compared the SHG from $La_4Ge_3S_{12}$ with that from a commonly used SHG material, KDP, using the same experiment setup. Our findings reveal that the coefficient of the quadratic term in $La_4Ge_3S_{12}$ ($0.602 \times 10^{-4}$/mW) is approximately half of that in KDP ($1.165 \times 10^{-4}$/mW).

Recently, ultraviolet NLO properties have been of increasing interest in a variety of applications., and were reported in a variety of materials including $CsSbF_2SO_4$ [24], $K_2Sb(P_2O_7)F$ [25], $Hg_3O_2SO_4$ [26], $Rb_2SbFP_2O_7$ [27], and $GeHPO_3$ [28], and a review of these materials is in Ref. [29]. For example, $CsSbF_2SO_4$ showed a strong SHG response of around 3.0 times that of KDP [24]. $La_4Ge_3S_{12}$ is also expected to show a larger SHG response by making it single crystalline and choosing the best phase-matching conditions.

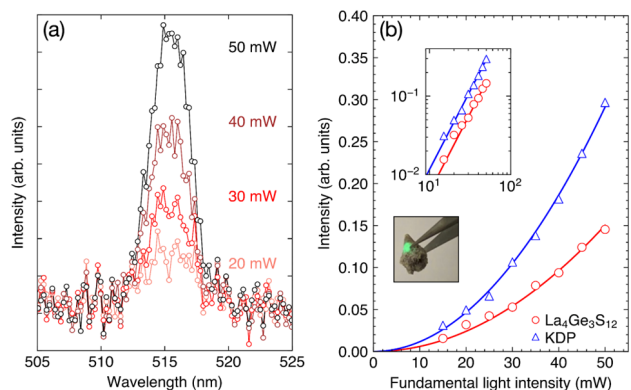

**Figure 6**. SHG from $La_4Ge_3S_{12}$ in response to an incident ultrashort pulsed laser (1030 nm). (a) SH intensity using different light intensities (20-50 mW). (b) SH intensity as a function of laser light intensity where the quadratic curves are used to fit the experimental data. The inset shows the same plot in the logarithmic scale. The inset photograph shows the visible SH green light (515 nm).

## Conclusions

In conclusion, $La_4Ge_3S_{12}$ was successfully grown using the direct reaction method. Powder and single-crystal XRD, EPMA, and XPS showed that the $La_4Ge_3S_{12}$ samples were polycrystalline with a single phase. The narrow scan XPS spectra revealed that the charge of the constituent elements La, Ge, and S were $(La^{3+})_4(Ge^{4+})_3(S^{2-})_{12}$ based on the peak positions of each element. Furthermore, our XPS results are in good agreement with DOS calculations for the valence-band. We also measured the NLO response in terms of the laser intensity dependence of SHG. We found that the SHG light emitted by $La_4Ge_3S_{12}$ was quadratically proportional to the laser intensity. The efficiency of the SHG from $La_4Ge_3S_{12}$ was found to be approximately half of that from KDP, a typical SHG compound.

## Experimental Section

**Synthesis**

To grow $La_4Ge_3S_{12}$, we used the following simple components for each element: 99.9% pure La pieces from Rare Metallic Co., Ltd., and 99.99% pure Ge and S powder from Kojundo Chemical Lab. Co., Ltd. The elements were sealed in an evacuated quartz tube with La: Ge:S = 4:3:12 ratio in a high vacuum of base pressure $10^{-3}$ Pa. The temperature treatment process of synthesis of the $La_4Ge_3S_{12}$ compound is as follows. The reagents were mixed and heated from 27 ℃ to 250 ℃ over 24 hours. The temperature was maintained at 250 ℃ for 2 hours, then increased to 950 ℃ over 20 hours. The mixture was held at 950 ℃ for 100 hours before being slowly cooled to 500 ℃ over 24 hours and quenched to room temperature. Following synthesis, the resulting compound showed a SHG response for 1 year after being kept in dry air (see Fig. 7 for SHG after 1 year.) We also confirmed that the sample was not damaged even under a 1 W irradiation: the maximum laser power used in the present work. Therefore, we did not confirm the clear onset of the damage threshold. Since the domains in polycrystalline materials are randomly distributed, it was not possible to satisfy the phase-matching conditions in our samples. We adopted the direct reaction method because the resulting polycrystalline samples are practical enough for SHG measurements in this study.

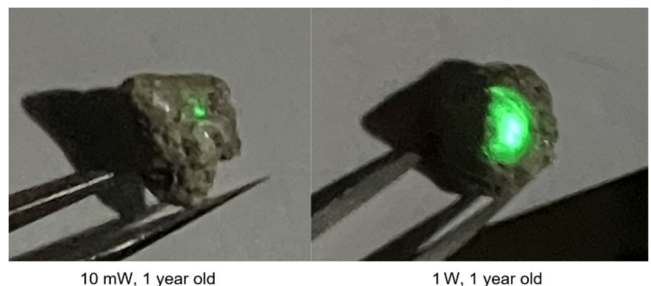

**Figure 7** SHG from $La_4Ge_3S_{12}$ after 1 year. (Left) Under 10 mW irradiation. (Right) Under 1 W irradiation.

**Powder and single-crystal X-ray diffraction.**

We performed powder X-ray diffraction with a RIGAKU Mini Flex II diffractometer at 30 kV/15 mA using Cu-Kα X-rays (λ = 1.5406 Å) in the angle range 5º ≤ 2Θ ≤ 100º. Lattice constants were determined by Rietveld analysis with RIETAN-FP [30]. In addition, we isolated a small single-crystal from the polycrystalline samples and carried out single-crystal XRD analysis on a Rigaku Rapid imaging plate diffractometer at 296 K with Mo-Kα radiation (λ = 0.71073 Å). The crystal structure was determined using the direct method with Sir2019 [31] and refined by the full-matrix least-squares method with the SHELXL-2019/2 software [32]. A multi-scan absorption correction was applied. The experimental settings are summarized in Table 1 together with crystallographic and refinement data. The above structural solution and refinements were performed using WinGX software [33].

**Electron probe microanalysis**

We determined the chemical composition of our samples using electron probe microanalysis (EPMA) with a JEOL FE-EPMA JXA-8530FPlus. Due to the insulating nature of $La_4Ge_3S_{12}$, the sample was carbon-coated before analysis of the sample to prevent charging effects during analysis. The energy of the incident electron beam was fixed at 20 kV.

**X-ray photoemission spectroscopy**

X-ray photoelectron spectroscopy (XPS) spectra were measured at room temperature using monochromatic Al-Kα X-ray's ($h\nu$ = 1486.6 eV) with a PHI 5000 VersaProbe III from ULVAC-PHI. The total energy resolution was about 500 meV. The samples studied were in powder form. To avoid charging effects during XPS analysis of the insulating $La_4Ge_3S_{12}$, we used a neutralizing gun.

**Second harmonic generation**

We measured the laser intensity dependence of SHG using the setup shown in Fig. 8 instead of employing the conventional Perry Kurtz method [34]. We observed SHG from the intensity dependence of reflected light using an ultrashort pulsed laser PHAROS with an infrared wavelength of 1030 nm. The pulse width is 200 fs, the repetition rate is 1 kHz, and the spot size is approximately 2 mm in diameter. The output power of the laser was swept from 15 mW to 50 mW. We crushed the samples into a powder form, pressurized them at 10 MPa, and pelletized them for SHG measurements. KDP was also ground into a fine powder and measured like $La_4Ge_3S_{12}$. The infrared light was directed onto the sample which was attached to a glass slide with cellophane tape. The reflected light passes through a short-pass filter (Thorlabs FESH0950 cut-off wavelength: 950 nm, damage threshold: 1.0 J/cm$^2$) to remove reflected infrared light, and we measured the resulting SHG intensity with a compact spectrometer (Thorlabs CCS200, Wavelength Range: 200 - 1000 nm). This setup allowed us to accurately measure the SHG intensity of $La_4Ge_3S_{12}$ and compare it with that of the reference material, providing valuable insights into the NLO properties of $La_4Ge_3S_{12}$.

# ARTICLE

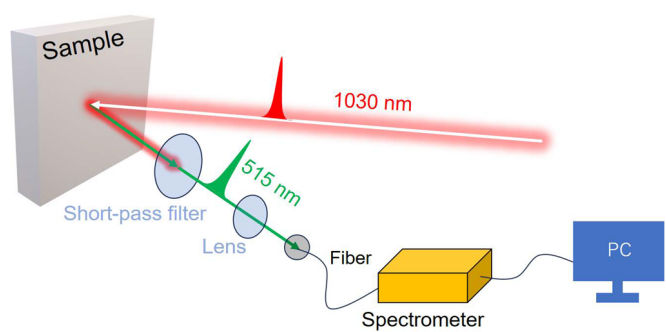

**Figure 8.** Experimental setup of the SHG measurements performed for the $La_4Ge_3S_{12}$ sample. The incident ultrashort infrared pulsed laser light (1030 nm) is generated from a PHAROS laser. A short-pass filter is used between the sample and detector to minimize the reflected fundamental light, resulting in a selective detection of the SHG by the spectrometer.

**First-principles calculations**

We theoretically evaluated the DOS of $La_4Ge_3S_{12}$ by using density functional theory calculations within the strongly constrained and appropriately normed (SCAN) meta-generalized gradient approximation (SCAN meta-GGA) [35,36]. SCAN meta-GGA simulations were performed using the Vienna Ab initio Simulation Package (VASP) [37-42]. The DOS calculations were based on the crystal structure data from a previous study [14]. In the VASP calculations, the crystal structure was replaced by an optimized structure based on the meta-GGA energy function. As a result, the experimental crystal structure was reproduced within a few percent.

## Acknowledgments

We thank N. Nakabayashi and P. D. Bentley for many fruitful discussions. This work was supported by MEXT Quantum Leap Flagship Program (MEXT Q-LEAP) Grant No. JPMXS0118068681 and JSPS KAKENHI Grant-in-Aid No. 19H05824.

**Keywords:** Second harmonic generation, electronic structure, $La_4Ge_3S_{12}$,

**Entry for the Table of Contents** (Please choose one layout)

Layout 1:

# FULL PAPER

Second harmonic generation from $La_4Ge_3S_{12}$ in response to the incident laser (1030 ± 10 nm). Fundamental light intensity dependence of the second harmonic intensity is shown. Curves are the fits using a quadratic function. The inset shows the same plot in the logarithmic scale. The photograph shows the second harmonic light visible in green (515 nm).

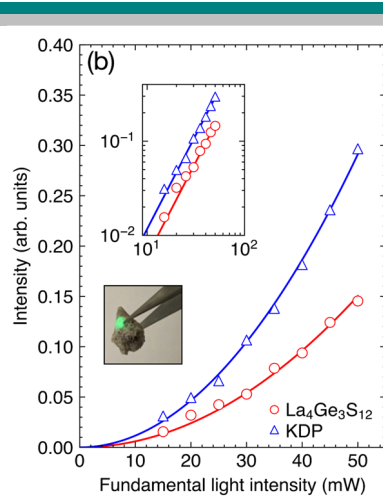

*Hiroya Ohtsuki, Suguru Nakata, Yu Yamane, Ryunosuke Takahashi, Koichi Kusakabe, and Hiroki Wadati\**

*Page No. – Page No.*

**Second harmonic generation of $La_4Ge_3S_{12}$**

Additional Author information for the electronic version of the article.

Suguru Nakata: 0000-0003-1675-9532
Yu Yamane: 0000-0002-5872-3063
Ryunosuke Takahashi: 0000-0002-6099-5201
Koichi Kusakabe: 0000-0002-0206-8864
Hiroki Wadati: 0000-0001-5969-8624